\documentclass[journal]{IEEEtran}
\usepackage{amsmath,graphicx,cite,epsfig}
\usepackage{multirow}
\usepackage{booktabs}
\usepackage{float}
\usepackage{subfigure}
\usepackage{caption}
\usepackage{threeparttable}
\usepackage{diagbox}
\usepackage{url}
\usepackage{bm}
\usepackage{stfloats}
\usepackage{amsfonts}

\ifCLASSINFOpdf

\else

\fi

\begin{document}
%
\title{Dehazed Image Quality Evaluation: From Partial Discrepancy to Blind Perception}
%
%
%

\author{Wei Zhou, \IEEEmembership{Member, IEEE}, Ruizeng Zhang, Leida Li, \IEEEmembership{Member, IEEE}, \\ Hantao Liu, \IEEEmembership{Senior Member, IEEE}, and Huiyan Chen

\thanks{This work was supported in part by NSFC under Grant 62171340. W. Zhou is with the Department of Electrical and Computer Engineering, University of Waterloo, Waterloo, ON N2L 3G1, Canada (e-mail: wei.zhou@uwaterloo.ca).}
\thanks{R. Zhang and H. Chen are with Beijing Institute of Technology, Beijing 100081, China (e-mail: hireason@163.com; chen\_h\_y@263.net).}
\thanks{L. Li is with the School of Artificial Intelligence, Xidian University, Xi’an 710071, China. (e-mail: ldli@xidian.edu.cn).}
\thanks{H. Liu is with the School of Computer Science and Informatics, Cardiff University, Cardiff, CF24 4AX, United Kingdom (e-mail: liuh35@cardiff.ac.uk).}}

\maketitle

\begin{abstract}
Image dehazing aims to restore spatial details from hazy images. There have emerged a number of image dehazing algorithms, designed to increase the visibility of those hazy images. However, much less work has been focused on evaluating the visual quality of dehazed images. In this paper, we propose a Reduced-Reference dehazed image quality evaluation approach based on Partial Discrepancy (RRPD) and then extend it to a No-Reference quality assessment metric with Blind Perception (NRBP). Specifically, inspired by the hierarchical characteristics of the human perceiving dehazed images, we introduce three groups of features: luminance discrimination, color appearance, and overall naturalness. In the proposed RRPD, the combined distance between a set of sender and receiver features is adopted to quantify the perceptually dehazed image quality. By integrating global and local channels from dehazed images, the RRPD is converted to NRBP which does not rely on any information from the references. Extensive experiment results on several dehazed image quality databases demonstrate that our proposed methods outperform state-of-the-art full-reference, reduced-reference, and no-reference quality assessment models. Furthermore, we show that the proposed dehazed image quality evaluation methods can be effectively applied to tune parameters for potential image dehazing algorithms.
\end{abstract}


%
\IEEEpeerreviewmaketitle

\section{Introduction}
%
%
%
%

\IEEEPARstart{T}{he} visibility of images is vital for delivered consumers. Nowadays, we usually use visible light imaging equipment to acquire visual scenes. In such cases, when the outdoor environment is captured, image visibility would be inevitably degraded by possible bad weather conditions due to the scattering or absorption of light by atmospheric particles \cite{li2018benchmarking}. Among them, haze is one of the representative atmospheric phenomena. To reduce the influence of visibility damage caused by hazy effects, many image dehazing algorithms \cite{meng2013efficient,lai2015single,berman2016non} have been proposed and achieved great success in the field of vision-related systems.

\begin{figure}[t]
	\centerline{\includegraphics[width=9.1cm]{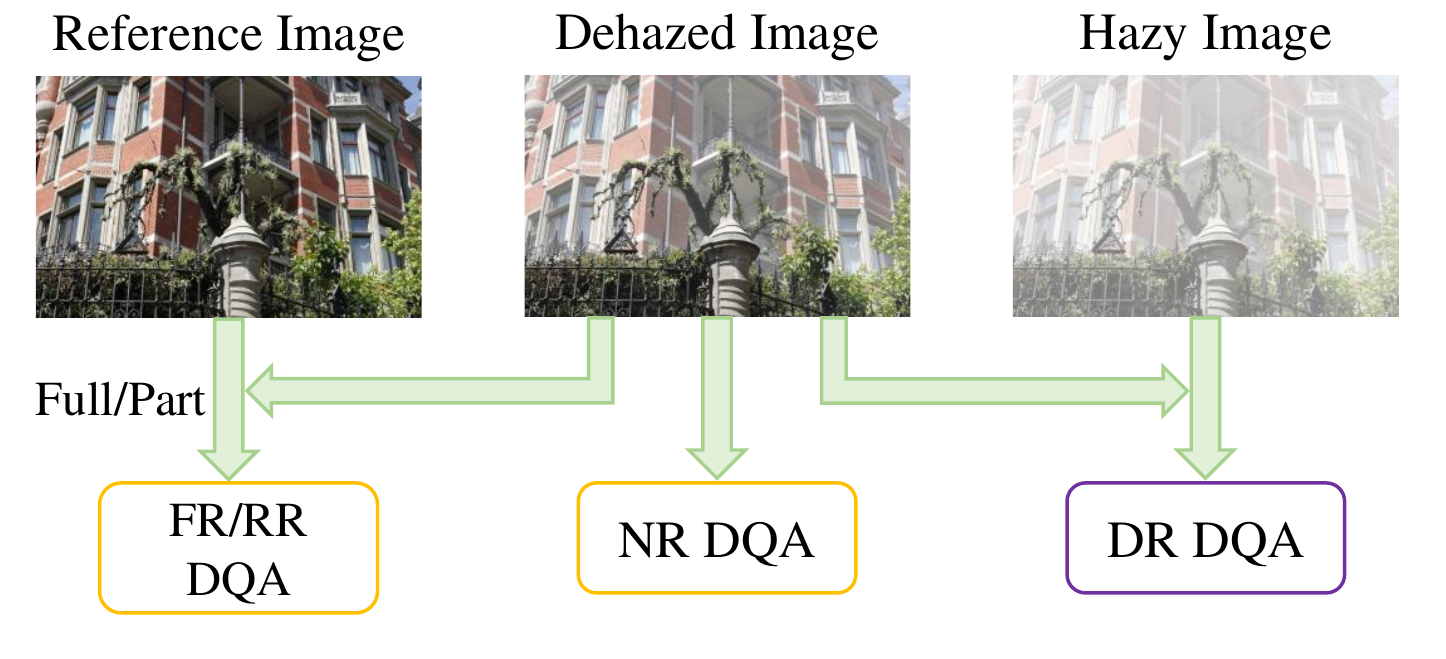}}
	\caption{Illustration of objective DQA categories, where FR and RR DQA methods use full and part reference information, respectively. The DR DQA models rely on hazy images, while NR DQA directly evaluates the perceptual quality from dehazed images.}
	\centering
	\label{figure1}
\end{figure}

With the development of image dehazing, dehazed images with a variety of appearances are generated by different image dehazing models. One nature question that arisen here is how to effectively evaluate and compare the performance of various image dehazing approaches? Because human subjects are the ultimate receivers of dehazed images, subjective quality evaluation \cite{ma2015perceptual} is the most accurate and reliable way to measure the visual quality of dehazed images. Moreover, based on subjective testing, several benchmarking dehazed image quality databases \cite{min2019quality,zhao2020dehazing,liu2020image} have been built, including regular and aerial scenarios. Nevertheless, subjective experiments are often time and labor consuming. The objective quality evaluation \cite{min2018objective} is an alternative to subjective ones, in which particularly objective metrics are designed to automatically estimate the perceptually dehazed image quality.

As shown in Fig. \ref{figure1}, similar to traditional image quality assessment (IQA), for objective dehazed quality assessment (DQA), there are generally four categories regarding to existing DQA metrics. These consist of full-reference (FR), reduced-reference (RR), degraded-reference (DR), and no-reference (NR) DQA models. Among them, the FR DQA supposes the pristine reference image is totally available, while the RR DQA only relies on part of the corresponding reference information. Since the hazy image is the degraded counterpart of the original reference image, based on this, the DR DQA methods can be developed. When both haze-free and hazy images are unavailable, we may resort to the NR DQA models which directly predict the visual quality from dehazed images.

Although lots of objective quality evaluation methods have shown effectiveness in estimating human perceived quality for conventional distortions such as noise \cite{shen2011hybrid}, JPEG compression \cite{golestaneh2013no}, blur \cite{li2019blind}, etc, the DQA models specifically designed for dehazed images are still in their infancy. As for FR and DR DQA situations, either original haze-free or degraded hazy images can be fully accessible, which is beneficial to the algorithm design of dehazed quality evaluation. However, in practical real-world applications, we may not obtain full information of them. In other words, it becomes more challenging for RR and NR DQA tasks that we focus on. Consequently, in this paper, we first propose a novel RR DQA method and then extend it to a more practical NR DQA model that does not rely on any other information except for the dehazed images. To the best of our knowledge, we are the first to propose a RR DQA method and have such an extension to the NR manner. Besides, our proposed quality evaluation models can be applied to tune parameters for image dehazing algorithms.

In our proposed frameworks, to evaluate the perceptual quality of dehazed images produced by various image dehazing algorithms, we consider the cognitive mechanisms of the human visual system (HVS). To be specific, when subjects perceive dehazed images, the luminance discrimination and color appearance are observed from low-level dimensions. Then, the high-level overall naturalness is also taken into account. This process is more likely to be reflected by the hierarchical properties of the human brain. Additionally, the HVS tends to aggregate the visual quality impression from both global and local channels.

\begin{figure*}[t]
	\centerline{\includegraphics[width=16.7cm]{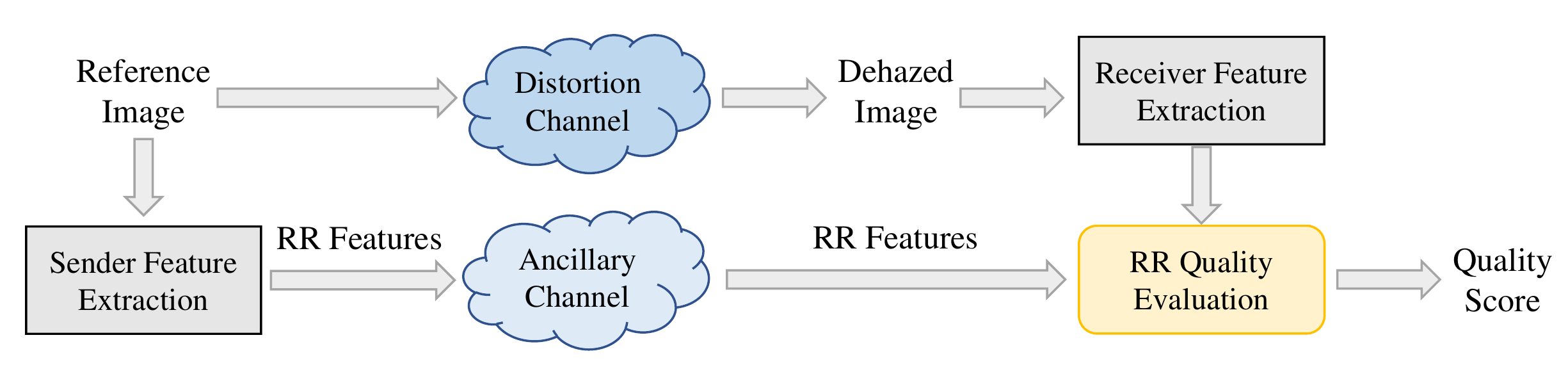}}
	\caption{Deployment of RR quality evaluation systems.}
	\centering
	\label{figure2}
\end{figure*}

Around the above-mentioned aspects, this paper makes the following main contributions:

\begin{itemize}
\item We propose RR and NR dehazed quality evaluation methods via partial discrepancy and blind perception, respectively.
\item Motivated by the hierarchical properties of dehazed image perceiving process, we propose quality-aware features, including luminance discrimination, color appearance, and overall naturalness.
\item Based on the characteristics of the HVS, the global and local aggregation is taken into consideration. Extensive experiments on dehazed image quality databases verify that our proposed metrics have superior performance. In addition, we optimize potential parameter-based image dehazing algorithms by applying our proposed dehazed image quality evaluation methods.
\end{itemize}

The rest of this paper is organized as follows. The related work of our paper is presented in Section II. Section III introduces the proposed Reduced-Reference dehazed image quality evaluation approach based on Partial Discrepancy (RRPD). In Section IV, we then propose the No-Reference quality assessment metric with Blind Perception (NRBP), which combines both global and local channels. We present experimental results and analysis in Section V, and finally conclude the work in Section VI.

\section{Related Work}
Objective IQA designs computational models to predict the human perception-based image quality, which is also known as mean opinion score (MOS). The simplest FR IQA model is peak signal-to-noise ratio (PSNR) that compares the pixel difference between reference and distorted images. However, such a signal fidelity model does not consider the characteristics of the HVS. Therefore, the PSNR cannot estimate the accurate visual quality as human observers perceive. To overcome the drawbacks of signal fidelity, Wang et al. proposed the well-known structural similarity (SSIM) index \cite{wang2004image} based on the visual perception of image structures. Besides, several variants of similarity measurement were also developed, such as the multiscale SSIM (MS-SSIM) index \cite{wang2003multiscale}, the information content weighted SSIM (IW-SSIM) measure \cite{wang2010information}, the feature similarity (FSIM) index \cite{zhang2011fsim}, the gradient similarity (GSM) model \cite{liu2011image}, the gradient magnitude similarity deviation (GMSD) \cite{xue2013gradient}, the perceptual similarity (PSIM) measure \cite{gu2017fast}, and the superpixel-based similarity (SPSIM) index \cite{sun2018spsim}, etc. Other FR IQA methods called information fidelity criterion (IFC) \cite{sheikh2005information} and visual information fidelity (VIF) \cite{sheikh2006image} evaluate visual quality from the aspect of image information.

\begin{figure*}[t]
	\centerline{\includegraphics[width=17.7cm]{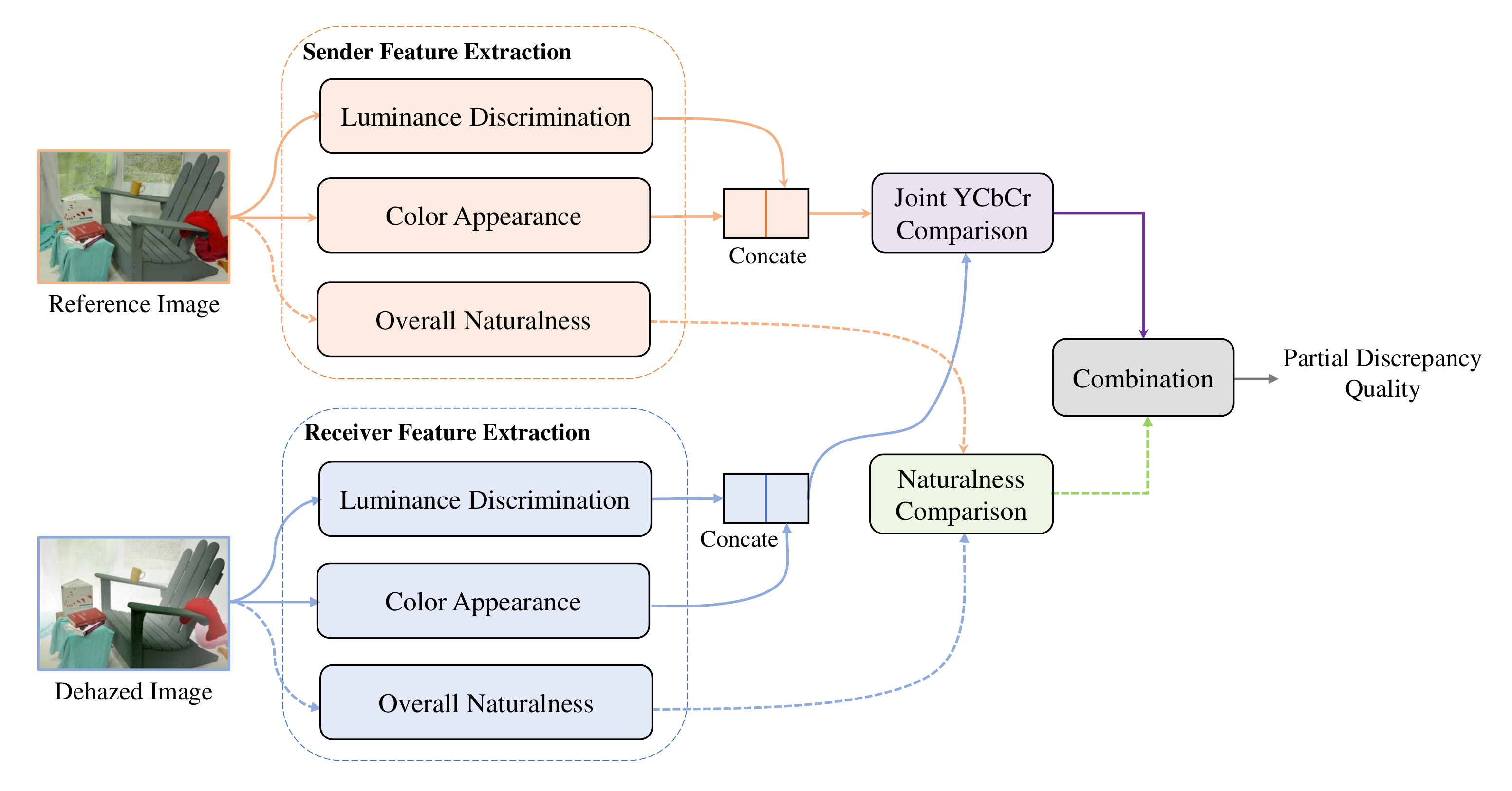}}
	\caption{Framework of the proposed RR quality evaluation method, where luminance discrimination, color appearance and overall naturalness are employed for feature extraction.}
	\centering
	\label{figure3}
\end{figure*}

Apart from the FR IQA approaches, many NR IQA methods have been proposed during the past decades. According to a spatial natural scene statistic (NSS) model, the classical blind/referenceless image spatial quality evaluator (BRISQUE) was built \cite{mittal2012no}. Other NSS models were also presented, e.g., the blind image quality index (BIQI) \cite{moorthy2010two}, the natural image quality evaluator (NIQE) \cite{{mittal2012making}}, the blind image integrity notator using DCT statistics (BLIINDS-II) \cite{saad2012blind}, the distortion identification-based image verity, and the integrity evaluation (DIIVINE) index \cite{moorthy2011blind}, etc. In \cite{wu2015highly}, a highly efficient NR IQA model named local pattern statistics index (LPSI) was presented to evaluate the image quality. The dipIQ index was proposed in \cite{ma2017dipiq}, where quality-discriminable image pairs were obtained to conduct learning-to-rank for perceptual quality assessment. In \cite{ma2017end}, Ma et al. designed a multi-task end-to-end optimized deep neural network (MEON) for NR IQA, which consists of a distortion identification network and a quality prediction network.

The RR IQA is a kind of objective way between FR IQA and NR IQA. The development of RR IQA falls behind and few models have been proposed for this category. For example, the reduced-reference image quality assessment method (RRIQA)  \cite{wang2005reduced} was proposed according to an image statistic model in the wavelet transform domain. Afterwards, based on the internal generative mechanism and visual saliency detection, the RQMSH \cite{wang2016reduced} and SIRR \cite{min2018saliency} were proposed to predict the distorted image quality with side information from reference data.

These traditionally generic IQA models introduced above are mainly designed for the quality assessment of natural images. However, the properties of natural images differ from that of dehazed images, even the artifacts generated by various image dehazing algorithms are unlike the traditional distortions such as common Gaussian white noise in natural images. Thus, it is necessary to devise specially dehazed quality evaluation methods. In the literature, representative FR DQA methods include the DEHAZEfr \cite{min2019quality} and FRFSIM  \cite{liu2020image}. Based on visibility and distortion measurement, the NR DQA metric named VDA-DQA \cite{guan2022visibility} was developed by complex contourlet transform. Due to the possible hazy images, one can also propose DR DQA models, e.g., the objective dehazing quality index (DHQI) \cite{min2018objective}.

Considering that none of the above attempts focus on the RR DQA problem, in this work, we aim to fill in this blank. Specifically, we propose a partial discrepancy based RR DQA model and then extend it to the NR DQA method with blind perception. Furthermore, our dehazed quality evaluation metrics can also be applied to optimize the parameter selection for the potential image dehazing algorithm.

\begin{figure}[t]
	\centerline{\includegraphics[width=9.1cm]{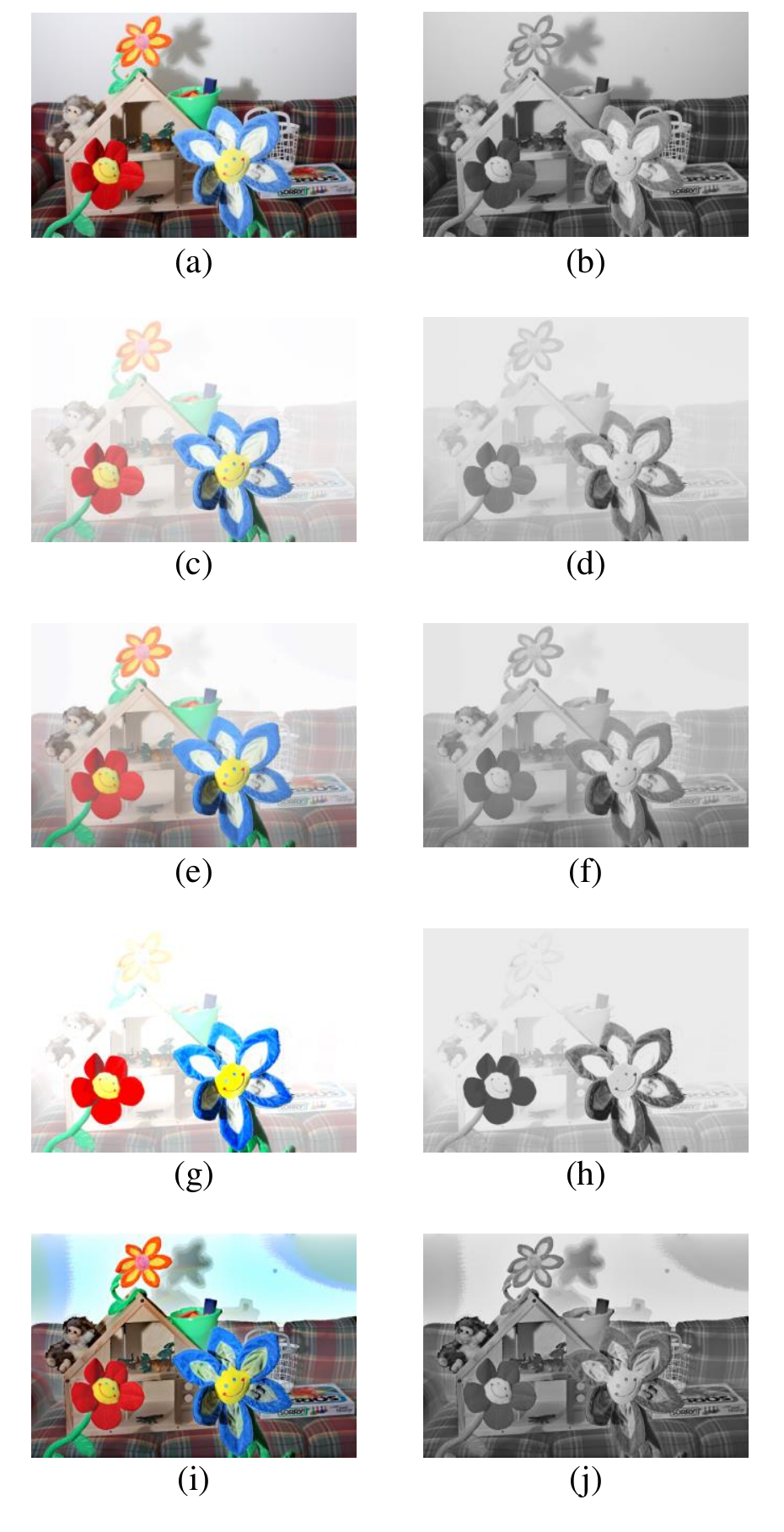}}
	\caption{Luminance quality degradation from image dehazing. (a) A haze-free image; (c) The corresponding hazy image; (e) Dehazed image with haze removal degradation; (g) Dehazed image with structural recovery degradation; (i) Dehazed image with enhancement effects degradation; (b), (d), (f), (h) and (j) are the luminance maps for (a), (c), (e), (g) and (i) respectively.}
	\centering
	\label{figure4}
\end{figure}

\section{Proposed RR Quality Evaluation Method}
In Fig. \ref{figure2}, we give the deployment of RR quality evaluation systems. According to this deployment, there are a feature extraction process at the sender side and a feature extraction followed by a quality evaluation process at the receiver side, for the reference and dehazed images, respectively. The dehazed image is generated through the distortion channel. Moreover, the RR features obtained from sender feature extraction usually have a much lower data rate than the original image data and are transmitted to the receiver by an ancillary channel.

The framework of our proposed RR quality evaluation method for dehazed images, i.e., RRPD, is shown in Fig. \ref{figure3}. In the proposed RR DQA method, motivated by the hierarchically dehazed image perceiving process, we first develop three groups of quality-aware features, containing luminance discrimination, color appearance, and overall naturalness. Then, the feature comparison and combination are used for partial discrepancy quality estimation. In the following subsections, we will introduce the technical details of RRPD.

\subsection{Luminance Discrimination}
When human perceives the dehazed images generated by different image dehazing algorithms, luminance information significantly influences perceptual image quality \cite{li2016blind}. Fig. \ref{figure4} shows some examples of the typical luminance quality degradation from image dehazing, where the corresponding luminance maps of reference, hazy, and dehazed images are presented. Our observation is that the quality degradation in the aspect of luminance information mainly comes from haze removal, structural recovery, and enhancement effects. Therefore, we propose the luminance discrimination features described as follows.

\textit{1) Haze Removal:} The haze in the image causes visibility loss. A lot of mainstream image dehazing approaches have been proposed to remove the haze as much as possible, which can help to increase the visibility. Under the circumstances, more haze removal can reflect better visual quality. Here we employ several statistics in the luminance component to quantify the haze removal degrees. Let $R$ denote the luminance map of reference image. The resolution is $M\times N$. To reveal the haze removal, we first compute the mean and standard deviation of luminance map by:

\begin{equation}
\operatorname{MEA}=\frac{1}{M \times N} \sum_{m=1}^M \sum_{n=1}^N R(m, n),
\end{equation}

\begin{equation}
\operatorname{STD}=\sqrt{\frac{1}{M \times N} \sum_{m=1}^M \sum_{n=1}^N[R(m, n)-\overline{R}]^2},
\end{equation}
where $\overline{R}$ is the mean operation of luminance map, which is the same as Eqn (1).

Suppose the ordered list for the row of luminance map is $\widehat{R}$. The median value regarding to the row is computed as:

\begin{equation}
X=\frac{1}{2}\left(\widehat{R}_{\left\lfloor\frac{M+1}{2}\right\rfloor}+\widehat{R}_{\left\lceil\frac{M+1}{2}\right\rceil}\right).
\end{equation}
Then, the final median value of luminance map can be obtained by:

\begin{equation}
\operatorname{MED}=\frac{1}{2}\left(\widehat{X}_{\left\lfloor\frac{N+1}{2}\right\rfloor}+\widehat{X}_{\left\lceil\frac{N+1}{2}\right\rceil}\right),
\end{equation}
where $\widehat{X}$ represents the ordered list of $X$.
Similarly, we calculate the mode value for the row of luminance map as follows:

\begin{equation}
Y=\left\{R_i \mid P\left(R_i\right) \geq P\left(R_j\right), i \neq j\right\},
\end{equation}
where $P(\cdot)$ indicates the probability and the final mode value of luminance map is computed by:

\begin{equation}
\operatorname{MOD}=\left\{Y_i \mid P\left(Y_i\right) \geq P\left(Y_j\right), i \neq j\right\}.
\end{equation}

Except for the mean, standard deviation, median, and mode statistics, the global entropy that determines the `surprise’ of a specific image is also taken into consideration as:

\begin{equation}
\operatorname{ENT}=-\sum_{k=0}^K P_k \log \left(P_k\right),
\end{equation}
where $K$ is the maximum pixel value in the whole luminance map. Besides, $P_k$ represents the probability of pixel value equaling to $k$. With all the extracted luminance statistics, we can exploit them to form the quantification of haze removal degradation.

\begin{figure}[t]
	\centerline{\includegraphics[width=9.1cm]{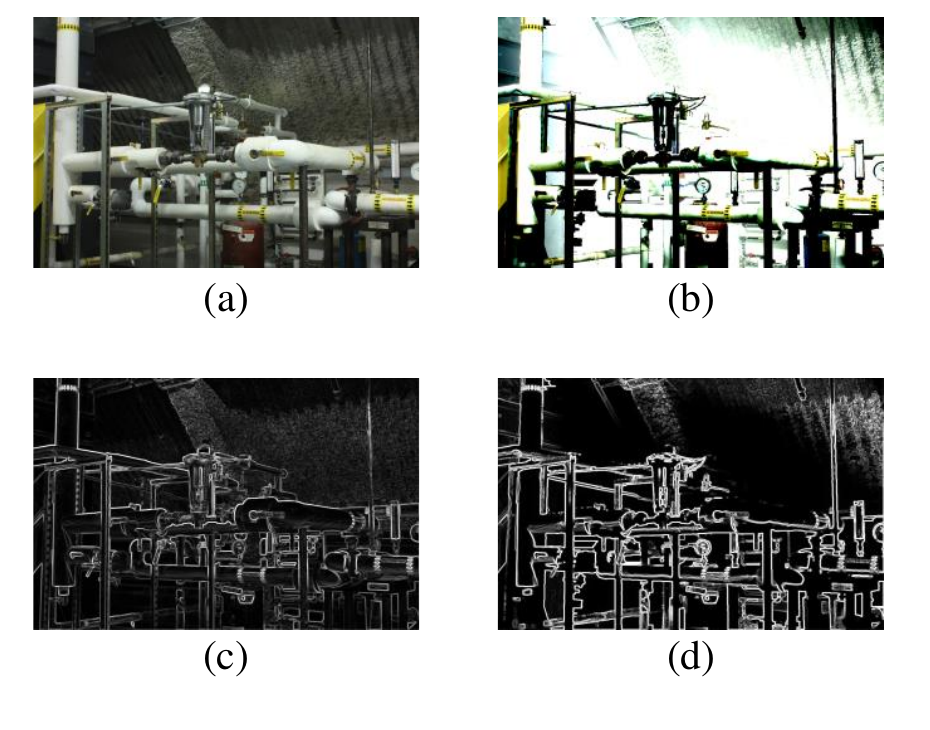}}
	\caption{Comparison of CSF weighted gradient maps. (a) A reference haze-free image; (b) Test dehazed image; (c),(d) The corresponding CSF weighted gradient map of (a) and (b).}
	\centering
	\label{figure5}
\end{figure}

\begin{figure*}[t]
	\centerline{\includegraphics[width=16.7cm]{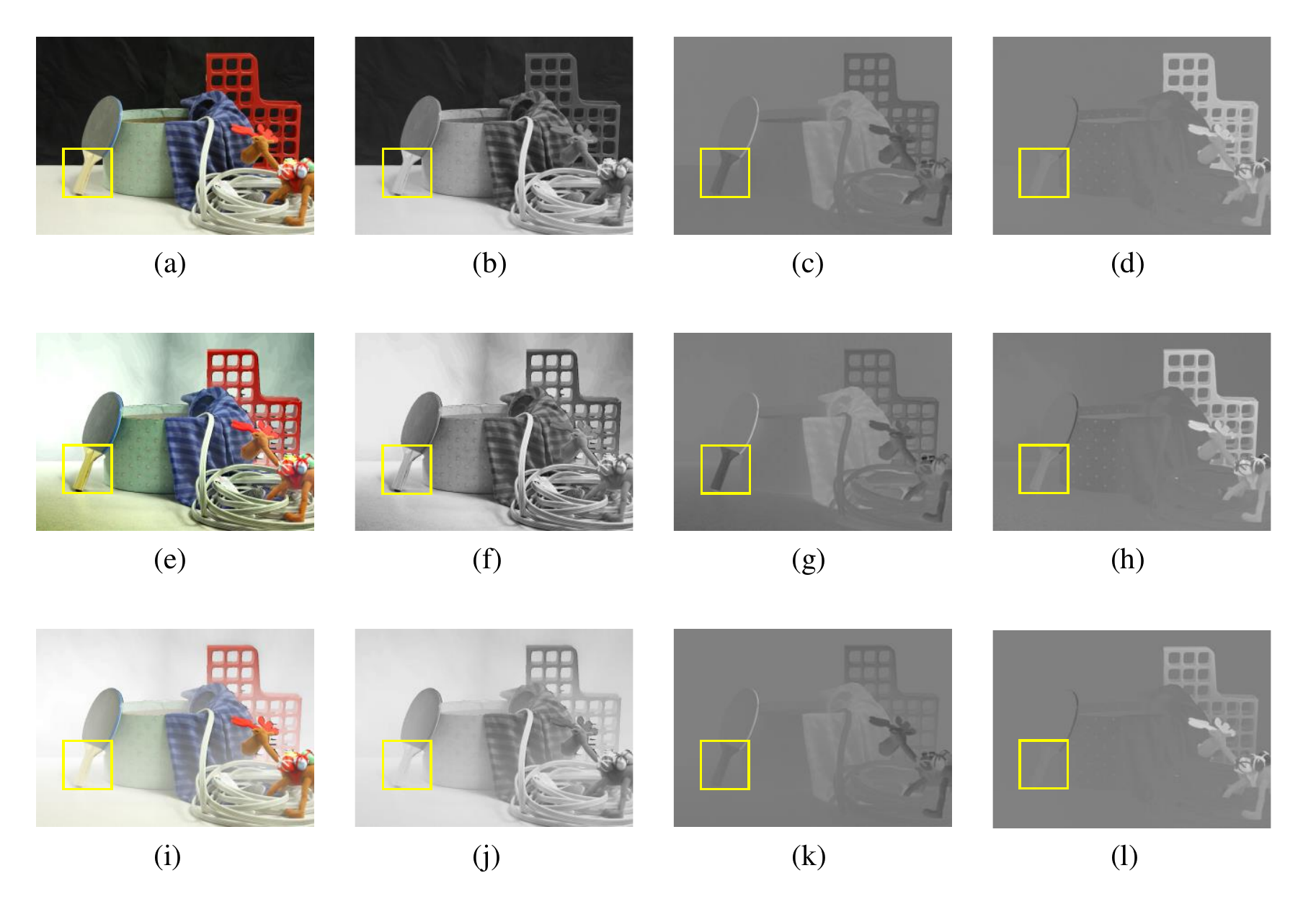}}
	\caption{Color decomposition results. (a) A reference haze-free image; (e) Dehazed image produced by Berman16 \cite{berman2016non}; (i) Dehazed image generated by Lai15 \cite{lai2015single}; (b-d) The corresponding YCbCr components of (a); (f-h) The corresponding YCbCr components of (e); (j-l) The corresponding YCbCr components of (i).}
	\centering
	\label{figure6}
\end{figure*}

\begin{figure*}[t]
	\centerline{\includegraphics[width=17.7cm]{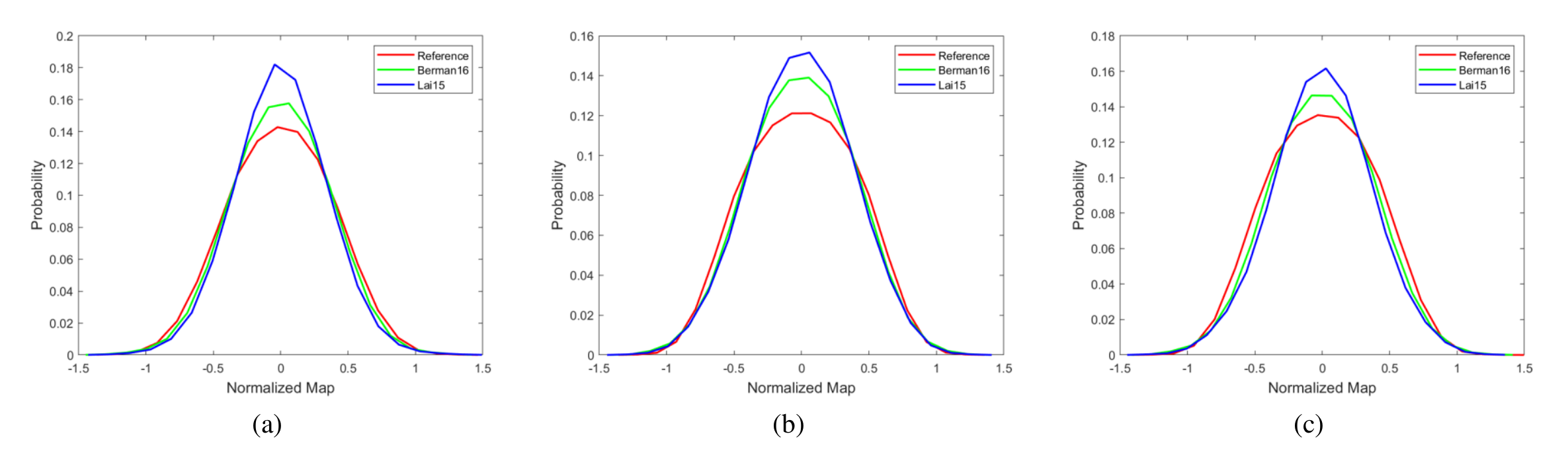}}
	\caption{Statistical distribution of different dehazed images from Fig. \ref{figure6} after normalization. (a) Statistical distribution for Luminance Y component; (b) Statistical distribution for chrominance Cb component; (c) Statistical distribution for chrominance Cr component.}
	\centering
	\label{figure7}
\end{figure*}

\textit{2) Structural Recovery:} Since structures are important for image quality evaluation, many IQA models based on image structures have been proposed \cite{wang2004image,min2016blind,gu2017evaluating,zhou2021image}. One main goal of image dehazing algorithms is also to recover the structures. From Fig. \ref{figure4} (g-h), we find that part of the structural information is missing in the dehazed images, which leads to the quality degradation. Thus, according to the characteristics of the HVS, i.e., the contrast sensitivity function (CSF) \cite{robson1966spatial}, we first filter the luminance map and then compute the gradient map weighting by CSF to reflect image structures.

Specifically, the CSF can measure the HVS sensitivity to various frequencies of visual stimulus. Based on CSF, human eyes have different abilities to discriminate target brightness at different spatial frequencies. Here we employ a modified CSF function \cite{golestaneh2016reduced} given by:

\begin{equation}
H(f, \varphi)=H_1(f, \varphi) H_2(f, \varphi),
\end{equation}
where $f$ denotes the radial spatial frequency with cycles per degree of visual angle (c/deg). Moreover, $\varphi \in[-\pi, \pi]$ represents the angular frequency. $H_1(f, \varphi)$ and $H_2(f, \varphi)$ are the frequency response from a circularly symmetric Gaussian filter and the frequency response of a CSF model originally proposed
in \cite{mannos1974effects} and adjusted in \cite{1987}. These can be computed as follows:

\begin{equation}
H_1(f, \varphi)=e^{-2 \pi^2 \alpha^2 f^2},
\end{equation}

\begin{equation}
H_2(f, \varphi)=\left\{\begin{array}{l}
2.6\left(0.0192+\eta f_{\varphi}\right) e^{-\eta f_{\varphi}}, f \geq f_{\text {peak }} \\
0.981, \text { otherwise }
\end{array}\right.,
\end{equation}
where $\alpha=0.5$ is adopted to control the filter cutoff. Inspired by \cite{mitsa1993evaluation}, we set $\eta=0.114$. $f_\varphi$ is used to represent direction-based correction of $f$ to reduce the contrast sensitivity along the diagonal, which can be calculated by:

\begin{equation}
f_\varphi=\frac{f}{0.15 \cos (4 \varphi)+0.85}.
\end{equation}

We exploit the modified CSF function $H(f,\varphi)$ to filter the luminance map. As shown in Fig. \ref{figure5}, we then obtain the CSF weighted gradient map, which can reflect the quality variation. Finally, the mean, standard deviation, median, mode, and entropy statistics of CSF weighted gradient map are adopted to measure the structural recovery degradation.

\begin{figure*}[t]
	\centerline{\includegraphics[width=16.7cm]{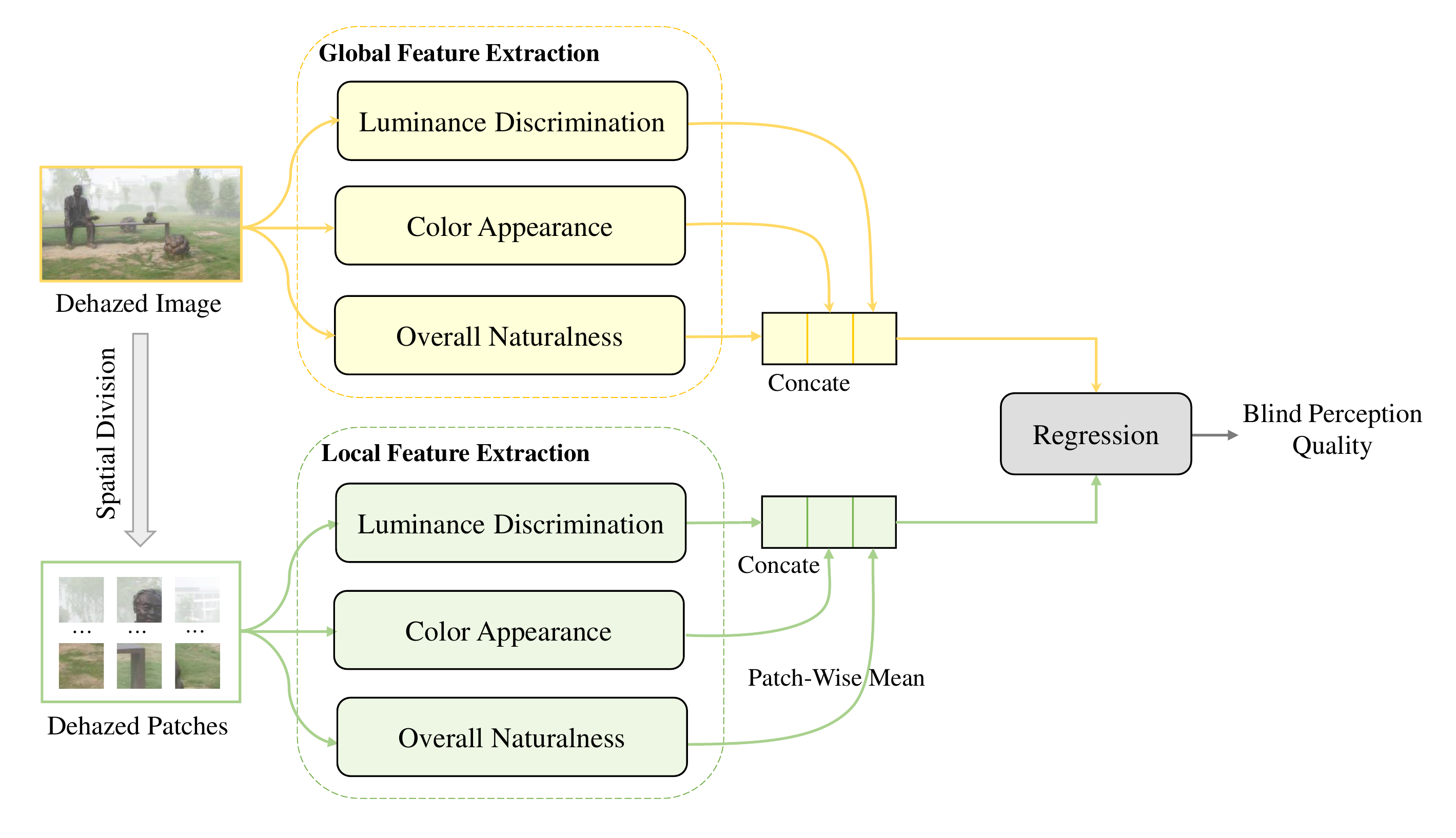}}
	\caption{Framework of the proposed NR quality evaluation method, where global and local channels are used for feature extraction.}
	\centering
	\label{figure8}
\end{figure*}

\textit{3) Enhancement Effects:} By using different image dehazing algorithms, the resulting dehazed images may be over-enhanced, as demonstrated by Fig. \ref{figure4} (i-j). In order to measure the contrast variation caused by image dehazing, we use the mean of local variance and normalized local variance \cite{min2018objective} which can determine the enhancement effects.

Specifically, the local variance of luminance map $R$ is computed by:

\begin{equation}
\sigma(x, y)=\sqrt{\sum_{u, v} w(u, v)[R(x+u, y+v)-\mu(x, y)]^2},
\end{equation}
where $x, y$ indicate the pixel indexes in the spatial domain. $w$ is a Gaussian weighting window. Moreover, $\mu$ represents the local mean, which can be calculated as:

\begin{equation}
\mu(x, y)=\sum_{u, v} w(u, v) R(x+u, y+v).
\end{equation}
Considering that the local variance is sensitive to the local mean, we also derive the normalized local variance as follows:

\begin{equation}
\gamma(x, y)=\frac{\sigma(x, y)}{\mu(x, y)+C},
\end{equation}
where $C$ represents a small positive constant used for avoiding instability. After obtaining the local variance and normalized local variance of luminance map, the mean values of them are employed to quantify the quality degradation from enhancement effects.

\subsection{Color Appearance}
Apart from luminance information, color cues are also useful to help the HVS perceive visual signals. There exist many works that have investigated the effects of luminance and chrominance on perceptual image quality \cite{lee2015towards,temel2016csv,lee2016toward,wang2022generation}. In these works, chrominance has been proven to achieve a promising gain for the image quality evaluation task. Therefore, we first convert the original image $I$ into YCbCr components by:

\begin{equation}
I \rightarrow \{R, C b, C r\},
\end{equation}
where $R$ is the reference luminance map. $Cb$ and $Cr$ are two chrominance maps.

We show the color decomposition results in Fig. \ref{figure6}, which include two image dehazing algorithms, i.e., Berman16 \cite{berman2016non} and Lai15 \cite{lai2015single}. From this figure, it can be seen the color differences in the reference and dehazed ones, indicating that color information can be applied to measure the dehazed image quality. Thus, we then compute the mean, standard deviation, median, mode, and entropy statistics of chrominance to form the color appearance features.

\subsection{Overall Naturalness}
Since the HVS works in a hierarchical manner, we utilize the higher level overall naturalness features of the YCbCr components to estimate the quality degradation from a natural sense. To be specific, we first realize the decorrelation of input image, e.g., luminance map $R$ with divisive normalization transform \cite{lyu2011dependency} as below:

\begin{equation}
\tilde{R}(x, y)=\frac{R(x, y)-\mu(x, y)}{\sigma(x, y)+C},
\end{equation}
where the local variance $\sigma$ and local mean $\mu$ are obtained from Eqn (12) and Eqn (13), respectively. $C$ is the same constant as Eqn (14). Apart from the decorrelation, we also apply the whitening filter to make the distribution more Gaussian like.

In Fig. \ref{figure7}, we show the statistical distribution for YCbCr components. We see that the behaviors of curves are different for various image dehazing algorithms. Therefore, we then use the generalized Gaussian distribution (GGD) and more general asymmetric generalized Gaussian distribution (AGGD) \cite{lasmar2009multiscale} to fit the statistical distribution. The AGGD with zero mean value mode is given as:

\begin{equation}
g\left(\tau ; \lambda, \rho_l, \rho_r\right)=\left\{\begin{array}{l}
\frac{\lambda}{\left(\varsigma_l+\varsigma_r\right) \Gamma\left(\frac{1}{\lambda}\right)} e^{-\left(\frac{-\tau}{\varsigma_l}\right)^{\lambda}}, \tau<0 \\
\frac{\lambda}{\left(\varsigma_l+\varsigma_r\right) \Gamma\left(\frac{1}{\lambda}\right)} e^{-\left(\frac{-\tau}{\varsigma_r}\right)^\lambda}, \text { otherwise }
\end{array}\right.,
\end{equation}
where

\begin{equation}
\varsigma_l=\rho_l \sqrt{\frac{\Gamma\left(\frac{1}{\lambda}\right)}{\Gamma\left(\frac{3}{\lambda}\right)}},
\end{equation}

\begin{equation}
\varsigma_r=\rho_r \sqrt{\frac{\Gamma\left(\frac{1}{\lambda}\right)}{\Gamma\left(\frac{3}{\lambda}\right)}},
\end{equation}
and $\lambda$ controls the distribution shape. $\Gamma$ represents the Gamma function. $\rho_l$ and $\rho_r$ are the scale parameters of left and right sides. The AGGD becomes GGD when $\rho_l=\rho_r$. Moreover, the skewness and kurtosis of GGD are used as complementary features. The shape and scale parameters are estimated by the moment-matching based approach \cite{sharifi1995estimation}. In addition, the parameters $(\delta, \lambda, \rho_l^2, \rho_r^2)$ of the best AGGD fit are achieved in which $\delta$ calculated by:

\begin{equation}
\delta=\left(\rho_r-\rho_l\right) \frac{\Gamma\left(\frac{2}{\lambda}\right)}{\Gamma\left(\frac{1}{\lambda}\right)}.
\end{equation}
With YCbCr components, we adopt two scales, containing the original image scale and a reduced resolution scale proposed in \cite{mittal2012no} to perform as the overall naturalness features.

\begin{table*}[ht]
\begin{center}
\captionsetup{justification=centering}
\caption{\textsc{Performance Comparisons of Various Reduced-Reference Quality Evaluation Methods on Two Dehazed Subjective Databases.}}
\label{table1}
\scalebox{1.2}{
\begin{tabular}{|c|cccc|cccc|}
\hline
\multicolumn{1}{|c|}{Databases} & \multicolumn{4}{c|}{SHRQR} & \multicolumn{4}{c|}{SHRQA} \\ \hline
Methods & SRCC & KRCC & PLCC & RMSE & SRCC & KRCC & PLCC & RMSE \\ \hline
RRIQA \cite{wang2005reduced} &0.3540 &0.2487 &0.4378 &12.4543 &0.6624 &0.4741 &0.6694 &12.0025 \\
RQMSH \cite{wang2016reduced} &0.2793 &0.1883 &0.2846 &13.2794 &0.6403 &0.4565 &0.6566 &12.1855 \\
SIRR \cite{min2018saliency} &0.4662 &0.3222 &0.5053 &11.9535 &0.5997 &0.4153 &0.6090 &12.8148 \\
RRPD (without CSF) &0.6562 &0.4857 &0.7126 &9.7185 &0.7241 &0.5358 & 0.6998 & 11.5414 \\
\textbf{Proposed RRPD} &\textbf{0.7450} &\textbf{0.5633} &\textbf{0.8355} &\textbf{7.6115} &\textbf{0.7608}	&\textbf{0.5715} &\textbf{0.7271} &\textbf{11.0921} \\ \hline
\end{tabular}}
\end{center}
\end{table*}

\begin{table*}[ht]
	\begin{center}
		\captionsetup{justification=centering}
		\caption{\textsc{Performance Comparisons of Different Quality Evaluation Methods on SHRQR Database.}}
		\label{table2}
		\setlength{\tabcolsep}{8mm}{
			\scalebox{1.1}{
	\begin{tabular}{|c|c|cccc|}
		\hline
		Types & Methods & SRCC & KRCC & PLCC & RMSE \\ \hline
		\multirow{11}{*}{FR IQA} & PSNR &0.5972 &0.4231 &0.6523 &10.4996 \\
		& SSIM \cite{wang2004image} &0.5627 &0.3991 &0.6225 &10.8412 \\
		& MS-SSIM \cite{wang2003multiscale} &0.5836 &0.4160 &0.6275 &10.7855 \\				
		& IW-SSIM \cite{wang2010information} &0.5657 &0.4031 &0.6246 &10.8175 \\			
        & FSIM \cite{zhang2011fsim} &0.6256 &0.4615 &0.7418 &9.2889 \\			
        & IFC \cite{sheikh2005information} &0.5549 &0.4068 &0.7354 &9.3873 \\			
        & VIF \cite{sheikh2006image} &0.6287 &0.4705 &0.7609 &8.9885 \\			
        & GSM \cite{liu2011image} &0.6029 &0.4364 &0.6946 &9.9654 \\			
        & GMSD \cite{xue2013gradient} &0.6157 &0.4531 &0.7364 &9.3722 \\
        & PSIM \cite{gu2017fast} &0.6238 &0.4580 &0.7580 &9.0350 \\
		& SPSIM \cite{sun2018spsim} &0.6553 &0.4914 &0.7454 &9.2348 \\ \hline
		\multirow{8}{*}{NR IQA} & BIQI \cite{moorthy2010two} &0.0277 &0.0235 &0.2444 &13.4323 \\
        & BRISQUE \cite{mittal2012no} &0.4196 &0.2964 &0.5767 &11.3165 \\
        & NIQE \cite{mittal2012making} &0.4029 &0.2843 &0.5920 &11.1639 \\
        & BLIINDS-II \cite{saad2012blind} &0.3400 &0.2353 &0.5374 &11.6821 \\
        & DIIVINE \cite{moorthy2011blind} &0.3712 &0.2535 &0.5016 &11.9840 \\
        & LPSI \cite{wu2015highly} &0.3363 &0.2355 &0.5718 &11.3643 \\			
        & dipIQ \cite{ma2017dipiq} &0.0417 &0.0291 &0.1699 &13.6509 \\
		& MEON \cite{ma2017end} &0.2220 &0.1445 &0.3111 &13.1652 \\ \hline
		\multirow{2}{*}{FR DQA} & DEHAZEfr \cite{min2019quality} &0.8292 &0.6430 &0.8675 &6.8912 \\
        & FRFSIM \cite{liu2020image} &0.5862 &0.4264 &0.7467 &9.2141 \\ \hline 			
		\multirow{1}{*}{DR DQA} & DHQI \cite{min2018objective} &0.4240 &0.3000 &0.6739 &10.2338 \\ \hline
		\multirow{2}{*}{NR DQA} & VDA-DQA \cite{guan2022visibility} &0.4846 &0.3421 &0.6160 &10.9119 \\
		& \textbf{Proposed NRBP} &\textbf{0.8556} &\textbf{0.6823} &\textbf{0.8992} &\textbf{5.9771} \\ \hline
	\end{tabular}}}
	\end{center}
\end{table*}

\subsection{Partial Discrepancy Quality Estimation}
In analogy to the feature extraction of reference images, identical operations can be applied to the corresponding dehazed images. Suppose that we have the luminance discrimination, color appearance, and overall naturalness features from reference and dehazed images. These features are denoted by $LD$/$LD’$, $CA$/$CA’$, and $ON$/$ON’$, for haze-free and dehazed images, respectively.

We first concatenate $LD$ and $CA$ as well as $LD’$ and $CA’$, leading to $LDCA$ and $LD’CA’$. Then, the joint YCbCr comparison and naturalness comparison are computed, which can be combined as the final partial discrepancy quality. Thus, we obtain the proposed RR quality estimation as follows:

\begin{equation}
Q_{RRPD}=\mathcal{M}(\left|L D C A-L D C A^{\prime}\right|) \times \mathcal{M}(\left|O N-O N^{\prime}\right|),
\end{equation}
Where LDCA/LDCA’ and ON/ON’ are 21-dimensional and 120-dimensional vectors, respectively. $\mathcal{M}$ denotes the mean operation.

\section{Proposed NR Quality Evaluation Method}
In practical applications, the part of reference information may not be always available. Therefore, we extend the proposed RR quality evaluation method to a blind/no-reference dehazed image quality evaluation model called NRBP. The framework of the proposed NR quality evaluation method is shown in Fig. \ref{figure8}, where the only input is the dehazed image, and both global and local channels are used for feature extraction.

\subsection{Local and Global Channels}
Inspired by the characteristics of the HVS, to exploit more information from dehazed images, global and local channels are involved in the proposed NRBP framework. Specifically, we conduct a spatial division for the input dehazed image. Here the patch size is set to $32\times 32$. More experimental results can be found in Section V-E.

Similar to the RRPD, we have luminance discrimination, color appearance, and overall naturalness features for both channels. Then, we concatenate the extracted features separately. It should be noted that for local feature extraction, before the concatenation, the patch-wise mean is computed for all patch features of each dehazed image. As for aerial images, the local channel would lead to marginal performance improvement, which is validated in the experimental part (i.e., Section V-C). Therefore, we test the regular images to select the patch size.

\subsection{Blind Perception Quality Estimation}
 Finally, with the concatenated features from global and local channels, we utilize the support vector regression (SVR) \cite{chang2011libsvm} for the blind perception quality generation as:

\begin{equation}
Q_{NRBP}=\mathcal{F}(LD',CA',ON',LD_{local}',CA_{local}',ON_{local}'),
\end{equation}
where $\mathcal{F}$ is the SVR operation, and the common parameters are adopted for mainstream image quality evaluation works \cite{narwaria2010objective,chen2017blind,zhou2020tensor}.

\section{Experiments and Analysis}
In this section, we will present the experimental results and analysis of our proposed RR DQA and NR DQA models. First, we provide a brief overview of the adopted databases and criteria in our experiments. Then, the comparisons between the proposed metrics and state-of-the-art methods are conducted. Moreover, for NRBP, we also report the ablation study of individual channels as well as each proposed feature and parameter test with different local patch sizes. Finally, we demonstrate that our proposed metrics perform well for authentic dehazed images and can also serve as a useful tool to be applied for image dehazing.

\begin{table*}[ht]
	\begin{center}
		\captionsetup{justification=centering}
		\caption{\textsc{Performance Comparisons of Different Quality Evaluation Methods on SHRQA Database.}}
		\label{table3}
		\setlength{\tabcolsep}{8mm}{
			\scalebox{1.1}{
	\begin{tabular}{|c|c|cccc|}
		\hline
		Types & Methods & SRCC & KRCC & PLCC & RMSE \\ \hline
		\multirow{11}{*}{FR IQA} & PSNR &0.8246 &0.6397 &0.8040 &9.6080 \\
		& SSIM \cite{wang2004image} &0.8207 &0.6267 &0.8166 &9.3252 \\
		& MS-SSIM \cite{wang2003multiscale} &0.7895 &0.5893 &0.7815 &10.0811 \\
		& IW-SSIM \cite{wang2010information} &0.7949 &0.5954 &0.7841 &10.0276 \\
        & FSIM \cite{zhang2011fsim} &0.7424 &0.5471 &0.7348 &10.9583 \\
        & IFC \cite{sheikh2005information} &0.5630 &0.3871 &0.6140 &12.7522 \\
        & VIF \cite{sheikh2006image} &0.7048 &0.5384 &0.7651 &10.4044 \\
        & GSM \cite{liu2011image} &0.7832 &0.5892 &0.7719 &10.2715 \\
        & GMSD \cite{xue2013gradient} &0.7103 &0.5243 &0.6984 &11.5633 \\
        & PSIM \cite{gu2017fast} &0.7593 &0.5755 &0.7338 &10.9760 \\
		& SPSIM \cite{sun2018spsim} &0.8408 &0.6418 &0.8348 &8.8960 \\ \hline
		\multirow{8}{*}{NR IQA} & BIQI \cite{moorthy2010two} &0.3710 &0.2442 &0.3869 &14.8984 \\
        & BRISQUE \cite{mittal2012no} &0.1527 &0.1023 &0.3216 &15.2986 \\
        & NIQE \cite{mittal2012making} &0.3634 &0.2470 &0.4342 &14.5547 \\
        & BLIINDS-II \cite{saad2012blind} &0.0895 &0.0611 &0.3305 &15.2491 \\
        & DIIVINE \cite{moorthy2011blind} &0.2903 &0.1915 &0.3130 &15.3449 \\		
        & LPSI \cite{wu2015highly} &0.4170 &0.2902 &0.4780 &14.1918 \\
        & dipIQ \cite{ma2017dipiq} &0.0707 &0.0471 &0.1187 &16.0427 \\		
		& MEON \cite{ma2017end} &0.0339 &0.0246 &0.1268 &16.0283 \\ \hline
		\multirow{3}{*}{FR DQA} & DEHAZEfr \cite{min2019quality} &0.8615 &0.6685 &0.8554 &8.3692 \\
        & DEHAZEfr+ \cite{min2019quality} &0.9028 &0.7219 &0.9017 &6.9855 \\
        & FRFSIM \cite{liu2020image} &0.8065 &0.6245 &0.7844 &10.0216 \\ \hline
		\multirow{1}{*}{DR DQA} & DHQI \cite{min2018objective} &0.5675 &0.4341 &0.5726 &13.2458 \\	\hline
		\multirow{2}{*}{NR DQA} & VDA-DQA \cite{guan2022visibility} &0.6662 &0.4816 &0.6728 &11.9529 \\			
		& \textbf{Proposed NRBP} &\textbf{0.9158} &\textbf{0.7606} &\textbf{0.9170} &\textbf{6.3036} \\ \hline			
	\end{tabular}}}
	\end{center}
\end{table*}

\begin{table*}[t]
\begin{center}
\captionsetup{justification=centering}
\caption{\textsc{Ablation Test of Individual Channels on Two Dehazed Subjective Databases.}}
\label{table4}
\scalebox{1.2}{
\begin{tabular}{|c|cccc|cccc|}
\hline
\multicolumn{1}{|c|}{Databases} & \multicolumn{4}{c|}{SHRQR} & \multicolumn{4}{c|}{SHRQA} \\ \hline
Methods & SRCC & KRCC & PLCC & RMSE & SRCC & KRCC & PLCC & RMSE \\ \hline
NRBP (only local) &0.8024 &0.6252 &0.8719 &6.7006 &0.9078 &0.7447 &0.9098 &6.6050 \\
NRBP (only global) &0.8193 &0.6401 &0.8765 &6.6110 &0.9113	&0.7500	&0.9140	&6.4002 \\		
\textbf{NRBP (local + global)} &\textbf{0.8556} &\textbf{0.6823} &\textbf{0.8992} &\textbf{5.9771} &\textbf{0.9158} &\textbf{0.7606} &\textbf{0.9170} &\textbf{6.3036} \\ \hline
\end{tabular}}
\end{center}
\end{table*}

\subsection{Databases and Criteria}
To verify the proposed RR and NR dehazed quality evaluation methods, we carry out experiments and analysis on both synthetic and authentic dehazed image quality databases, i.e., SHRQR, SHRQA databases \cite{min2019quality} and exBeDDE \cite{zhao2020dehazing} database.

The SHRQR database contains 45 original haze-free images and 360 dehazed images generated by eight typical image dehazing algorithms. Each haze-free image is related to a corresponding hazy image.

The SHRQA database is composed of 30 pristine haze-free images and the corresponding 30 hazy images. Moreover, there exist 240 dehazed images produced by eight image dehazing algorithms which are the same as that of the SHRQR database. Note that the image content in this database is aerial.

The exBeDDE database is a dehazed image quality database based on real hazy images. There exist 1,670 dehazed images produced by 10 typical image dehazing approaches.

For comparison criteria, we adopt four commonly-used metrics as follows:

1. Spearman Rank-order Correlation Coefficient (SRCC) is defined by:

\begin{equation}
\centering
SRCC = 1 - \frac{{6\sum\limits_{t = 1}^T {{d_t}^2} }}{{T({T^2} - 1)}},
\end{equation}
where $T$ denotes total image numbers of the used database. $d_t$ is the rank difference between the $t$-th image's subjective and objective evaluations.

2. Kendall Rank Correlation Coefficient (KRCC) can be computed as:

\begin{equation}
\centering
KRCC = \frac{{2({F_c} - {F_d})}}{{T(T - 1)}},
\end{equation}
where $F_c$ and $F_d$ are the numbers of concordant and discordant pairs on the database, respectively.

3. Pearson Linear Correlation Coefficient (PLCC) is calculated as follows:

\begin{equation}
\centering
PLCC = \frac{{\sum\limits_{t = 1}^T {({s_t} - \overline s )({o_t} - \overline o )} }}{{\sqrt {\sum\limits_{t = 1}^T {({s_t} - \overline s )} \sum\limits_{t = 1}^T {({o_t} - \overline o )} } }},
\end{equation}
where $s_t$ and $o_t$ represent the $t$-th subjective and mapped objective quality scores. ${\overline s }$ and ${\overline o }$ are the corresponding mean values of $s_i$ and $o_i$, respectively.

4. Root Mean Squared Error (RMSE) can be estimated by:

\begin{equation}
RMSE = \sqrt {\frac{{\sum\limits_{t = 1}^T {{{({s_t} - {o_t})}^2}} }}{T}}.
\end{equation}

The SRCC and KRCC are applied to test prediction monotonicity and the ordinal association between two measured quantities. Meanwhile, the PLCC and RMSE are used to measure prediction accuracy. Note that higher correlation coefficients and lower error mean better performance.

Additionally, before computing the PLCC and RMSE values of different objective quality evaluation approaches, a five-parameter logistic nonlinear fitting function \cite{rohaly2000video} is employed to map the predicted quality scores into a common scale as:

\begin{equation}
\rm{\epsilon(q) = {\beta _1}(\frac{1}{2} - \frac{1}{{1 + {e^{({\beta _2}(q - {\beta _3}))}}}}) + {\beta _4}q + {\beta _5}},
\end{equation}
where $\rm{({\beta _1}...{\beta _5})}$ represent five parameters to be fitted. $\rm{q}$ and $\rm{\epsilon(q)}$ are the raw objective score generated by quality assessment models and the regressed objective score after the nonlinear mapping, respectively.

\begin{figure*}[t]
	\centerline{\includegraphics[width=16.1cm]{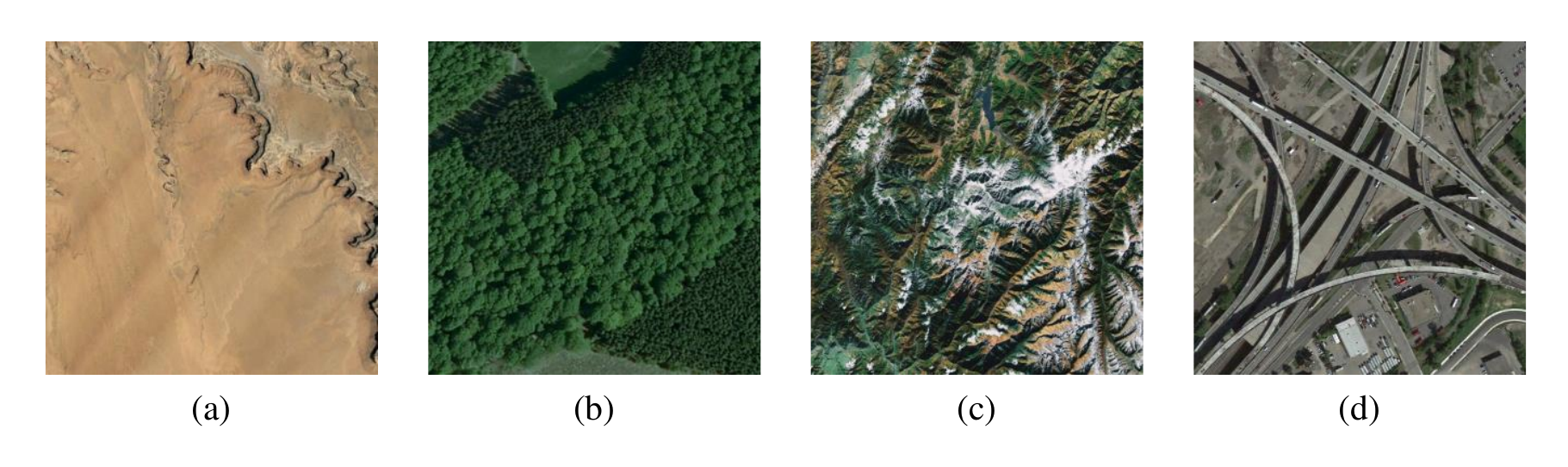}}
	\caption{Examples of aerial images. (a) Desert; (b) Forest; (c) Mountain; (d) Viaduct.}
	\centering
	\label{figure9}
\end{figure*}

\begin{table*}[t]
\begin{center}
\captionsetup{justification=centering}
\caption{\textsc{Performance Results of Each Proposed Feature (local + global) on SHRQR Database.}}
\label{table5}
\scalebox{1.2}{
\begin{tabular}{|c|c|c|c|c|}
\hline
Methods & SRCC & KRCC & PLCC & RMSE \\ \hline
Luminance Discrimination &0.6717 &0.5055 &0.8026 &8.1954 \\ \hline
Color Appearance &0.5002 &0.3592 &0.7141 &9.5582 \\ \hline
Overall Naturalness &0.8008 &0.6236 &0.8640 &6.8815 \\ \hline
\textbf{Proposed NRBP} &\textbf{0.8556} &\textbf{0.6823} &\textbf{0.8992} &\textbf{5.9771} \\ \hline
\end{tabular}}
\end{center}
\end{table*}

\begin{table*}[t]
\begin{center}
\captionsetup{justification=centering}
\caption{\textsc{Performance Results of Each Proposed Feature (local + global) on SHRQA Database.}}
\label{table6}
\scalebox{1.2}{
\begin{tabular}{|c|c|c|c|c|}
\hline
Methods & SRCC & KRCC & PLCC & RMSE \\ \hline
Luminance Discrimination &0.8214 &0.6454 &0.8220 &9.1890 \\ \hline
Color Appearance &0.6151 &0.4504 &0.6546 &12.0325 \\ \hline
Overall Naturalness &0.7327 &0.5461 &0.7495 &10.5121 \\ \hline
\textbf{Proposed NRBP} &\textbf{0.9158} &\textbf{0.7606} &\textbf{0.9170} &\textbf{6.3036} \\ \hline
\end{tabular}}
\end{center}
\end{table*}

\begin{table}[t]
\begin{center}
\captionsetup{justification=centering}
\caption{\textsc{Performance Variation of Different Patch Sizes for Proposed NRBP (local + global) on SHRQR Database.}}
\label{table7}
\scalebox{1.2}{
\begin{tabular}{|c|c|c|c|c|}
\hline
Patch Sizes & SRCC & KRCC & PLCC & RMSE \\ \hline
$16\times16$ &0.8482 &0.6729 &0.8971 &6.0858 \\ \hline
$32\times32$ &\textbf{0.8556} &\textbf{0.6823} &\textbf{0.8992} &\textbf{5.9771} \\ \hline
$64\times64$ &0.8527 &0.6808 &0.8981 &6.0351 \\ \hline
$128\times128$ &0.8506 &0.6776 &0.8930 &6.1905 \\ \hline
\end{tabular}}
\end{center}
\end{table}

\begin{table*}[ht]
	\begin{center}
		\captionsetup{justification=centering}
		\caption{\textsc{Performance Comparisons of Different Quality Evaluation Methods on exBeDDE Database.}}
		\label{table8}
		\setlength{\tabcolsep}{8mm}{
			\scalebox{1.1}{
	\begin{tabular}{|c|c|cccc|}
		\hline
		Types & Methods & SRCC & KRCC & PLCC & RMSE \\ \hline
		\multirow{8}{*}{NR IQA} & BIQI \cite{moorthy2010two} &0.3497 &0.2395 &0.3950 &0.2454 \\
        & BRISQUE \cite{mittal2012no} &0.3688 &0.2522 &0.4966 &0.2319 \\
        & NIQE \cite{mittal2012making} &0.1422 &0.0879 &0.2971 &0.2551 \\
        & BLIINDS-II \cite{saad2012blind} &0.3486 &0.2405 &0.3443 &0.2508 \\
        & DIIVINE \cite{moorthy2011blind} &0.4695 &0.3321 &0.5119 &0.2295 \\		
        & LPSI \cite{wu2015highly} &0.1212 &0.0918 &0.2417 &0.2592 \\
        & dipIQ \cite{ma2017dipiq} &0.3987 &0.2652 &0.4637 &0.2367 \\		
		& MEON \cite{ma2017end} &0.0227	&0.0150	&0.3604	&0.2492 \\ \hline
		\multirow{1}{*}{DR DQA} & DHQI \cite{min2018objective} &0.2717 &0.1825 &0.3058 &0.2555 \\	\hline
		\multirow{2}{*}{NR DQA} & VDA-DQA \cite{guan2022visibility} &0.4216	&0.2861	&0.4811	&0.2342\\
		& \textbf{Proposed NRBP} &\textbf{0.9115} &\textbf{0.7403} &\textbf{0.9354} &\textbf{0.0944} \\ \hline			
	\end{tabular}}}
	\end{center}
\end{table*}

\begin{figure*}[t]
	\centerline{\includegraphics[width=16.1cm]{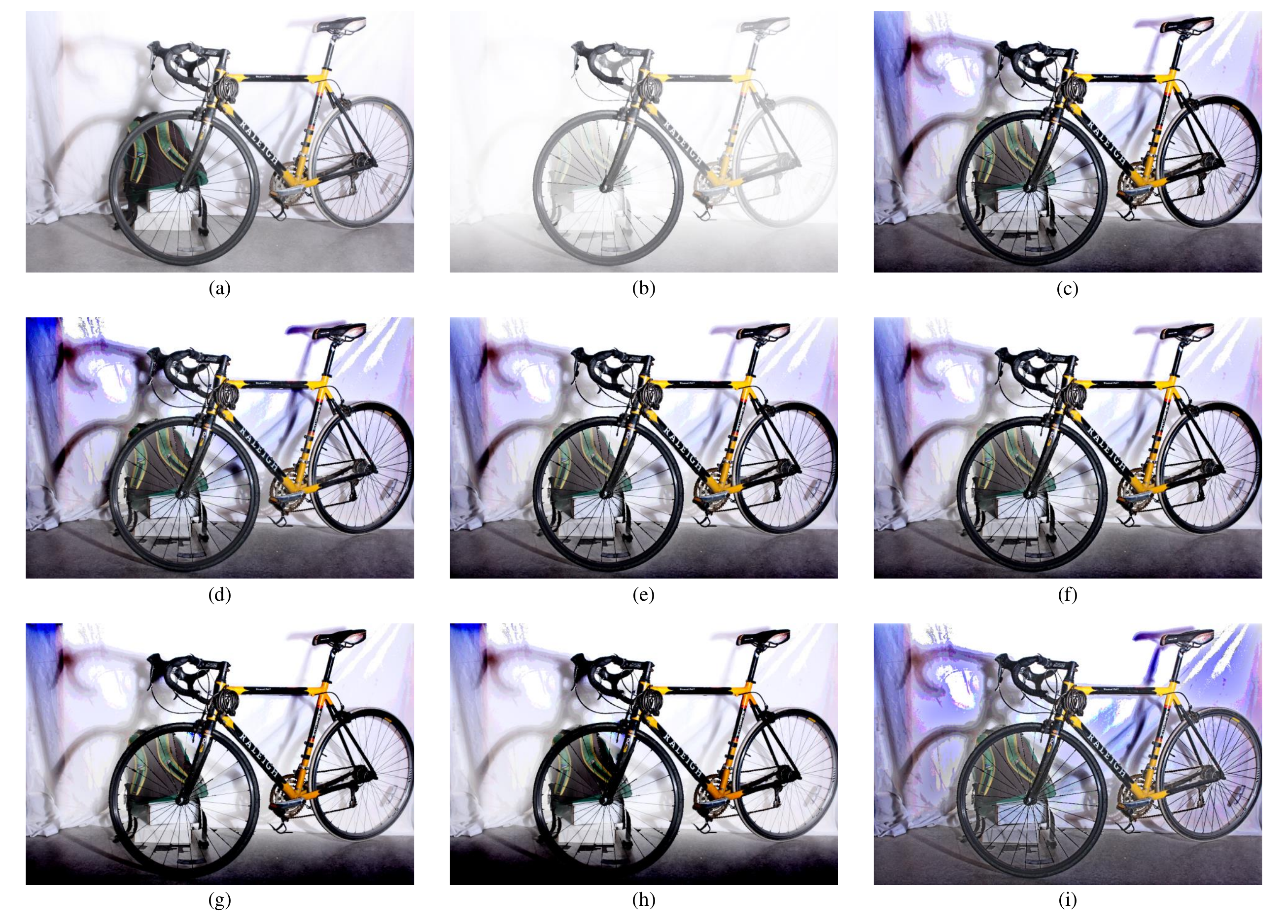}}
	\caption{Image dehazing by using different parameters. (a) A reference haze-free image; (b) The corresponding hazy image; (c) Dehazed image with default parameters, i.e., $\alpha=0.5$, $\lambda=1.0$ and $Q_{RRPD}=38.8803$; (d) Dehazed image with $\alpha=0.05$, $\lambda=1.0$ and $Q_{RRPD}=54.1494$; (e) Dehazed image with $\alpha=5$, $\lambda=1.0$ and $Q_{RRPD}=34.8126$; (f) Dehazed image with $\alpha=50$, $\lambda=1.0$ and $Q_{RRPD}=34.1356$; (g) Dehazed image with $\alpha=0.5$, $\lambda=0.1$ and $Q_{RRPD}=26.9847$; (h) Dehazed image with $\alpha=0.5$, $\lambda=0.01$ and $Q_{RRPD}=32.0869$; (i) Dehazed image with $\alpha=0.5$, $\lambda=10$ and $Q_{RRPD}=81.1083$.}
	\centering
	\label{figure10}
\end{figure*}

\subsection{Comparisons with Other Quality Evaluation Metrics}
We compare the proposed dehazed image quality evaluation methods with existing state-of-the-art ones. For the proposed RRPD, the other compared RR quality evaluation algorithms include RRIQA \cite{wang2005reduced}, RQMSH \cite{wang2016reduced}, and SIRR \cite{min2018saliency}. Besides, due to the lack of RR DQA metrics, we also test the RRPD without CSF. The performance results are reported in TABLE \ref{table1}.  From this table, we find that our specifically designed RR DQA model outperforms the others. Furthermore, removing CSF would deduce the performance of our model, which validates the effectiveness of the proposed CSF weighting. This is mainly attributed to the powerful perceptual properties of CSF, performing closer to the human visual perception in the dehazed image quality evaluation task.

As for the proposed NRBP, the traditional image quality evaluation methods for comparisons are 11 FR IQA and 8 NR IQA models. Except for these models, existing FR DQA, DR DQA and NR DQA approaches are also taken into consideration. The compared results on SHRQR and SHRQA databases are provided in TABLE \ref{table2} and TABLE \ref{table3}, respectively. As can be seen from these two tables, our proposed NRBP delivers the best performance on both databases. More importantly, the proposed metric even shows superior performance compared with DEHAZEfr \cite{min2019quality} and FRFSIM \cite{liu2020image} which are two FR DQA models. It should be noted that the DEHAZEfr+ is the improved DEHAZEfr measure for aerial images. Additionally, we notice that the performance numbers of the SHRQA database are generally better than that of the SHRQR database. This proves that dehazed quality evaluation for regular images is more challenging.

\subsection{Ablation Study of Individual Channels}
Since the NRBP architecture has global and local channels, the ablation study is adopted to test the results of individual channels. Specifically, we test three configurations, including the NRBP with only local channel, the NRBP with only global channel, and the NRBP with both local and global channels.

TABLE \ref{table4} provides the ablation results which are also verified by the significant t-test. We observe that the global channel performs better than the local channel and the combination of the two can boost the final results to some extent. However, the performance improvements for the SHRQR database are obviously larger than that for the SHRQA database. In Fig. \ref{figure9}, we show some examples of aerial images. An interesting trend is that aerial images are inclined to have similar textures. That is, the self-similarity of local patches for these images is relatively higher, compared to that of the regular images. This may be the main reason for that the proposed NRBP with only global channel obtains top-performing results on SHRQA database.


\subsection{Validity of Each Proposed Feature}
It is interesting to verify each proposed feature category regarding to luminance discrimination, color appearance, and overall naturalness.

In TABLE \ref{table5} and TABLE \ref{table6}, we provide the  performance results of individual proposed features on SHRQR and SHRQA databases, respectively. It can be seen that the proposed combination (i.e., NRBP) has the best performance, which demonstrates the superiority of our proposed metric. As for SHRQR and SHRQA databases, the overall naturalness and luminance discrimination outperform other individual proposed features separately. This may be due to the differential characteristics of regular and aerial images.

\subsection{Test with Different Local Patch Sizes}
Due to the relatively marginal improvement of adding local channel for aerial images, we test different local patch sizes on the SHRQR database.

As illustrated in TABLE \ref{table7}, several patch sizes are adopted in the experiments, including $16\times16$, $32\times32$, $64\times64$, and $128\times128$. From the results, we find that the $32\times32$ leads to the best performance. In addition, with the increase and decrease around $32\times32$, the performance would drop. This is also consistent with other visual quality evaluation methods \cite{kang2014convolutional,zhou2020blind,zhang2021deep}. Therefore, we choose the patch size equaling to $32\times32$ in our designed NRBP model.

\subsection{Evaluation on Authentic Dehazed Images}
In order to further verify the proposed NR quality evaluation method, we adopt an authentic dehazed image quality database (i.e., exBeDDE database).

We show the test results in Table \ref{table8}. Since the hazy images are available, here we compare the proposed NRBP with state-of-the-art NR IQA, DR DQA, and NR DQA models. We find that existing traditional NR IQA metrics generally fail to predict the dehazed image quality. In contrast, our proposed NRBP significantly outperforms the NR IQA models. Additionally, the proposed NRBP can also exceed the DHQI \cite{min2018objective} and VDA-DQA \cite{guan2022visibility} which are DR DQA and NR DQA methods specifically designed for dehazed images.

\subsection{Application to Image Dehazing}
A good dehazed quality indicator can not only predict the perceptually visual quality of dehazed images, but also effectively optimize the existing image dehazing algorithms. Here we also validate our proposed dehazed quality evaluation method by applying it to parameter selection of a classical image dehazing approach \cite{meng2013efficient}.

We give image dehazing results in Fig. \ref{figure10}. Since the proposed RRPD does not need the training process, here we show the objective quality scores obtained by RRPD. In this figure, higher $Q_{RRPD}$ indicates worse visual quality. The image dehazing algorithm has two parameters, including $\alpha$ and $\lambda$. Among the generated results, (c) utilizes the original parameters proposed in \cite{meng2013efficient}. We observe that the proposed NRBP can efficiently select better parameters. That is, (g) with $\alpha=0.5$, $\lambda=0.1$ delivers the best perceptual quality.

\section{Conclusion}
In this paper, we have presented a new RR dehazed quality evaluation method to predict the visual quality of dehazed images with only part of the original reference information. In the proposed framework, to involve the hierarchical property of the human perception, we propose quality-aware features from the aspects of luminance discrimination, color appearance, and overall naturalness. Then, motivated by the characteristics of the HVS, we extend it to a blind/NR quality evaluation model by integrating both global and local channels. Finally, experimental results on both synthetic and authentic dehazed image quality databases demonstrate the superiority of our proposed quality evaluation methods. Additionally, our proposed quality metric can also be applied to optimize the existing image dehazing algorithm by tuning its parameters.

In the future, we plan to explore more potential applications of the proposed dehazed quality evaluation methods. For example, using the metric as a loss function to train a more robust neural network for image dehazing. Besides, developing quality evaluation methods for video dehazing and its optimization is also a promising research direction.

\bibliographystyle{IEEEtran}
\bibliography{references}

%

%
%
%




\end{document}